\documentstyle[aps,preprint,eqsecnum]{revtex}
\newcommand{\be}{\begin{equation}}
\newcommand{\ee}{\end{equation}}
\newcommand{\bea}{\begin{eqnarray}}
\newcommand{\eea}{\end{eqnarray}}
\newcommand{\ra}{\rightarrow}
\newcommand{\wed}{\wedge}
\newcommand{\del}{\delta}
\newcommand{\tr}{ {\rm Tr }\,}
\newcommand{\real}{{\rm I \! R}}
\begin{document}
\preprint{\vbox{\hbox{SU-4240-647}\hbox{September, 1996}
\hbox{hep-th/9609226}}}
\title{ Edge Dynamics for BF Theories and Gravity }
\author{ Arshad Momen }
\address{Department of Physics, Syracuse University\\
Syracuse, NY 13244-1130, U.S.A.}
\maketitle
\begin{abstract}
We discuss BF theories defined on manifolds with
spatial boundaries. Variational arguments show that one needs to augment 
the usual action with a boundary term for specific 
types of boundary conditions. We also show how to use this procedure to 
find the
boundary actions for theories of gravity with first order formulations.
Possible connection with the membrane approach is also discussed.
\end{abstract}
\section{Introduction}

Gauge field theories defined on 
 manifolds with boundaries support observables defined on the 
boundaries called edge states \cite{bal1}.
Their existence can be established from very general arguments and for 
certain cases they can 
generate interesting current algebras\cite{witten}. These edge variables
represent degrees of freedom living on the boundary and can capture 
a lot of interesting physical information. For example,
for condensed matter systems like Quantum Hall samples\cite{wen,chandar}
or certain types of superconductors \cite{paolo}, a lot of interesting
phenomena happen on the boundary, and these edge states provide us with a 
natural framework to describe the system, when one's attention is focused 
on the boundary. It is important to appreciate
in this connection that the nature of  
the edge degrees of freedom is not completely arbitrary
but is rather constrainted by the bulk dynamics and therefore
can reflect the properties of the bulk.

Recently, we managed to establish the existence of such states also 
in gravity, when one excises a certain part of spacetime, for 
instance, a black hole\cite{us}. These states then have support 
on the boundary of the spacetime region that has been cut out.
Like their condensed matter analog, they can give important 
clues to our understanding of black hole physics, for 
instance, the origin of black hole entropy\cite{carlip}. In this 
connection, in  
a  previous paper we managed to show in a toy model in (2+1) dimensions 
 that the entanglement entropy \cite{rafael} 
one gets after tracing out the edge degrees of freedom scales with the 
perimeter ( the ``area law") in the weak coupling limit\cite{us2}. This 
result definitely encourages one
 to look for a similar situation in four dimensions
which might be of relevance for black hole physics.

Given the existence of these variables, an important puzzle is 
the dynamics of these edge variables in term of an action on the 
boundary.
The solution to this question is partially known 
for certain systems, e.g.
three dimensional gauge theories which  
include the Chern-Simons term \cite{Frolich,chandar}. 
In this  paper, we will show that similar 
results can be obtained 
for field theories which contain 
the so-called topological
BF term. This will shed light on the problem of constructing 
the action for gravitational edge states.

The pure BF
theory\cite{horowitz,blau} is known to be a topological
field theory. 
Unlike the Chern-Simon action which can be defined only in three 
dimensions,
one can define the BF action in  
any spacetime dimension. This action has  
appeared in various forms in different contexts,
being responsible for novel 
statistics for stringy objects \cite{kauffman} or the theory of 
Josephson junction arrays\cite{sodano}  as well as
for the 
classification of 2-knots \cite{cattaneo}. The form of the BF
action is very much 
similar to the first order formulations of gravity like the Palatini
formalism \cite{peldan}
or the self-dual ( or Ashtekar ) formalism \cite{sam}. 
In fact, one particular motivation for this investigation was to find 
the dynamics for the edge states appearing in gravity\cite{us}.
The
results obtained here can  be 
applied to gravity theories, specially to  
two-form gravity  theories \cite{CDJ}.

The paper is arranged as follows. 
In section 2, we review how the presence 
of the boundary leads us to a boundary action for the edge variables where 
the bulk dynamics is governed by a 
Chern-Simons theory. In 
section 3, we will follow similar steps to construct a boundary action 
for the abelian BF theory defined on 3+1 dimensional manifolds. In section 4,
we extend our results to the non-abelian case.  Next, in section 
5, we show how to  
write a boundary action for gravity in one of its 
first order formulations.
Finally, We discuss the
possible uses of the boundary action.

We will be using the differential  
form notation extensively in this paper 
though at times we will use 
the component notation also.

\section{Chern-Simons action with Boundary}
For orientational purposes, in this section
we will describe the abelian case only. However,
the extension to the 
non-abelian case is fairly straightforward. 
The treatment given below is also quite standard in effective
theories 
of the Quantum Hall Effect \cite{callan,wilczek,chandar}.

Let our spatial slices be diffeomorphic
to a disk $D$, so that the spacetime 
${\cal M}$ is $ D \times \real$. The Chern-Simons action defined on 
this three-dimensional manifold which has a boundary $\partial {\cal M }
= S^1 \times \real$ 
is given by
\be
S_0 = \frac{k}{4\pi}\int_{\cal M} A \wed dA. 
\label{1.0}
\ee

Because of the boundary, this action is not
invariant under a gauge transformation
\be
 A \rightarrow A + d \Lambda ,
\label{1.1}
\ee
as it picks up the
surface term
\be
S_{0} \rightarrow S_0 -\frac{k}{4\pi}\int _{\partial {\cal M}}d \Lambda \wed A,
 \label{1.2}
\ee
$\partial {\cal M}$ being $\partial D\times \real = S^1 \times \real$.
If we require that the system  be described by a gauge invariant action,
then we must add a surface term
to the Chern-Simons action $S_0$ to get an action
$S$ as follows:
\begin{eqnarray}
&&S_{tot} =S_0+\frac{k}{4\pi}\int _{\partial {\cal M}}d\phi \wedge A 
-\alpha \int _{\partial {\cal
M}} D \phi \wed *D \phi , \label{1.3}\\
&&D \phi \equiv d\phi - A. \label{1.4}
\end{eqnarray}
Under the gauge transformation,
(\ref{1.1}), $\phi$ simultaneously transforms as
\be
\phi \rightarrow \phi +  \Lambda ,\label{1.5}
\ee
rendering (\ref{1.3}) gauge invariant.

One can think of $\phi$ as the phase of a complex Higgs field, $\Phi
 =\rho e^{i\phi}$, where the amplitude $\rho$ is frozen to 
$\rho=v$. 

Note that we have added a
kinetic energy term for the field $\phi$.
The parameter  $\alpha$ is a constant. 
This term is not needed for gauge invariance of (\ref{1.3}), but gives
interesting dynamics to $\phi$.

So far, the coefficient $\alpha$ in (\ref{1.3}) has remained
 arbitrary.  However, if this
action were to describe QHE \cite{chandar}, then the currents due to the edge
scalar field $\phi$ are required to be ``chiral''\cite{wilczek}.  
For instance, if we demand that the currents are  left-moving, the condition 
one should impose  is
\be
D \phi -* D \phi =0. \label{1.7}
\ee
This fixes the coefficient $\alpha$ to be 
\be
\alpha = \frac{k}{8 \pi}.
\label{1.8}
\ee
Thus the correct  action describing QHE is
\be
S_{tot} =S_{bulk} + \frac{k}{4\pi}\int _{\partial {\cal M}}
d \phi\wed  A-\frac{k}{8\pi}\int _{\partial {\cal
M}}D \phi \wed *D \phi . 
\label{1.9}
\ee

We will show below that analogous actions can be constructed for the 
BF systems as well. 

It is interesting to note that boundary actions, similar to
the above have been constructed for the non-abelian Chern-Simons case by 
variational arguments \cite{carlip2} and gauge arguments\cite{us} also.

\section{Abelian BF theory}

We will now discuss how to write boundary actions for BF type field
theories defined on manifolds with spatial boundaries. The constraint
analysis for such 
field theories is discussed in \cite{paolo} where it is  
shown that they too contain
edge states \cite{paolo}. In contrast to these papers, which focus on the
canonical approach,
we will stress variational arguments here. They can be 
applied to the case of gravity too.

Let us  begin with a 3+1 dimensional spacetime 
${\cal M} \equiv 
 \real  \times \Sigma$ where the spatial part
$\Sigma$ is a manifold with the boundary  $\partial \Sigma$. So, the 
boundary of the spacetime is  $\partial {\cal M}  
\equiv \real \times  \partial \Sigma$.
The action describing the BF system is 
\be
S_0= \int_{\cal M} B \wed F= 
\frac{1}{2}\int_{\real \times \Sigma}dt d^3 x \epsilon^{ijkl} B_{ij}\partial_k A_l,
\label{2.1}
\ee
where $B= \frac{1}{2}B_{ij} dx^i\,dx^j$ is a two-form and 
$F=dA= \partial_j A_k dx^j dx^k$ is the  curvature form constructed 
out of the connection form  $A$.

The equation of motion for $A$ can be obtained by varying the $B$ field in 
the bulk and is 
\be
F= 0.
\label{2.2}
\ee

Notice that the equation of motion and the action as well
are trivially invariant under 
the gauge transformation
\be 
A \ra A + d \Psi 
\label{2.2a}
\ee
where $\Psi$ is a zero-form.

To get the equation of motion for $B$ one has to vary $A$. Under a 
generic variation of $A$,
the variation of the action is given by
\be
\del S_0 = \int_{\cal M} [ B \wed d(\del A)] = \int_{\cal M}
 [ - dB \wed \del A ] + \int_{\partial {\cal M}} [ B \wed \del A].
\label{2.3}
\ee
The sign for the boundary term is fixed by the orientation in
the bulk : $\epsilon^{t\,ijk} = \epsilon^{ijk}$ where
 $\epsilon^{123}=1$.

So, there is a surface term in $\delta S$ for a generic variation of $A$.
We wish to have such a
surface term equal to zero. 
This can be achieved in any one of the following ways: 
\begin{enumerate}
\item The variation $\del A$ is chosen to
vanish at the boundary,
\be
\del A|_{\partial {\cal M}}=0.
\label{2.4a}
\ee
\item One can choose $B$ to be zero on the boundary,
\be
B|_{\partial {\cal M}} = 0.
\label{2.4b}
\ee

\item One also has the option of adding surface terms. 
\end{enumerate}

The first two options are no good when 
nontrivial edge dynamics are sought for.
 As for the last option, one can 
 augment the action $S_0$ with the surface 
term 
\be
S'= - \int_{\partial {\cal M} } B \wed A.
\label{2.4}
\ee

 The option we would be exploring now for the augmented action $S_0 + S'$
is to vary $B$ in the bulk {\it without}
changing its boundary value which can be {\it nonzero} :
\bea
\del B |_{\partial {\cal M}}= 0, \nonumber \\ 
B|_{\partial{\cal M}} = \mbox{ Not necessarily zero.}\label{2.4c}
\eea
This is  different from (\ref{2.4b}). 
The choice of this type of boundary condition can be motivated from 
 the fact that $A$ is a connection whereas $B$ is  not.

When one looks at 
the augmented action 
\bea
S &=& S_0 + S' = \int_{\cal M}
 B \wed F  - \int_{\partial {\cal M}} B \wed A \nonumber \\
&=& \int_{\cal M} A \wed d B, 
\label{2.6}
\eea
one sees that this is nothing but the London action
\be
S= - \int_{\cal M} *j \wed A,
\label{2.7}
\ee
with  the dual identification $j = * dB$ of the 
current \cite{paolo}. But, when one has other terms in the
action which involve the coupling of the $B$ field to other fields,
evidently
such a simple interpretation is not necessarily available.
 
The equation of  motion for $B$  following from (\ref{2.6}) is
\be
dB = 0.
\label{2.8}
\ee

The action  (\ref{2.6}) now enjoys the gauge invariance
\be
B \ra B + d \chi,
\label{2.9}
\ee
where $\chi$ is a one-form. 

It is interesting that the action (\ref{2.6})
can also be constructed by demanding invariance of the action under the 
transformation (\ref{2.9}) alone without using the 
boundary conditions on $B$.  
  
 The modified
action   (\ref{2.6}) has lost its invariance under
the usual gauge  transformation
\be
A \ra A + d \Lambda.
\label{2.9a}
\ee
To remedy this situation,
 we introduce a scalar
field $\Phi$ on the edge ( i.e. on the boundary ) transforming as
\be
\Phi \ra \Phi - \Lambda,
\label{2.10}
\ee
so that the combined one-form 
\be
A'\equiv A + d \Phi = D \Phi 
\label{2.10a}
\ee
remains  invariant under the gauge transformation.
Given these quantities, we can write down the following gauge 
invariant action  inspired by our 
previous experience with the Chern-Simon case :
\bea
 S_{tot} &=&\int_{\cal M} [ B \wed F] - \int_{\partial {\cal M}} 
[ B \wed A'  
 + \lambda ( A' \wed * A') ] \nonumber \\
& \equiv &\int_{\cal M} B \wed F + S_{boundary}.
\label{2.11}
\eea
Here the Hodge $*$ operation is defined with respect to the 
boundary metric and $\lambda$ is a constant with  mass 
dimension 1. 

We would like to stress that as this action was found by fixing the 
boundary value of $B$,
we are not allowed to make a variation 
of $B$ on the
boundary to get the equations of motion on the boundary. 
However, one is allowed to vary $A$ or $\Phi$ at the boundary. 
Varying $\Phi$ one gets the equation of motion for 
$\Phi$ on the boundary:
\be
d (\lambda * D \Phi - B )=0
\label{2.12}
\ee
while varying $A$ one gets 
\be
\frac{ \del S}{\del A}=0 \Rightarrow \lambda 
D \Phi =0
\label{2.13}
\ee
Compatibility with the equation (\ref{2.13}) then 
lead to the condition 
\be
d B = 0 \Rightarrow B = d \xi
\ee
that is on the boundary $B$ is to be a ``pure gauge''.

The total action (\ref{2.11})  can also be motivated from the boundary
theory. This can be seen as follows. A charged scalar 
field $\Phi$ coupled to the antisymmetric field $B$ is described by 
\be
S_{boundary} = \int_{\partial M} [\lambda (d \Phi + A) ^*(d \Phi + A) +  
(d \Phi + A) \wedge B ]
\label{2.14}
\ee
However, 
varying $B$ in above equation 
we get the  condition 
\be
D\Phi = 0
\label{2.17}
\ee
on the boundary. On the other hand varying $A$ we are led to 
\be
\frac{1}{2}*\lambda ( D \Phi) +  B =0
\label{2.18}
\ee 
It must therefore 
be that there exists another term such that the term involving 
the variation of $A$ on the boundary involving 
the $B$ field exactly cancels out. This, as
can easily be seen, turns out to be the BF term defined in the bulk. 
Hence, the 
action (\ref{2.11}) hangs  together neatly.  This picture is basically 
the same as the one where one can argue for the existence of the Chern-Simons
action on a disk starting from the edge action for a gauged
chiral scalar field 
theory \cite{callan,chandar}.

One can easily show that
the above picture is  
applicable to arbitrary spacetime dimensions.

 \section{Non-Abelian BF system}

Given the above construction  for edge dynamics for 
 the abelian BF system, one can generalize  
it to non-abelian BF systems also. For the non-abelian case,
the two-form $B$ and the connection
$A$ are valued in a Lie algebra $G$. We also assume 
that this Lie algebra is endowed with an invariant trace, which 
we denote by $\tr$.
The BF action is then described by
\be
S_0 =  \int_{\cal M} \tr B \wed F
\label{4.1} 
\ee
where  the curvature form $F$ is now given by
\be
F= dA + A \wed A
\label{4.2}
\ee

The equation of motion arising from the variation of $B$ in the bulk is
\be
F = 0
\label{4.3}
\ee
However, a surface term like the one in 
 (\ref{2.3}) appears when one tries to vary $A$. Therefore, as
before we have to add a boundary action 
\be
S_b =  - \int_{\partial {\cal M}} \tr B \wed A
\label{4.4}
\ee
to allow for a non-zero boundary value for $B$.

We assume that the gauge transformation 
for $A$ now takes the standard form of the non-abelian connection :
\be
A \ra  g A g^{-1}  +g dg^{-1},
\label{4.5}
\ee
while $B$ transforms as 
\be
B \ra g B g^{_1}.
\label{4.5a}
\ee
Note that $S_0+ S_b$ is not invariant under (\ref{4.5}). So,
we introduce an edge field transforming as 
\be
h \ra g h 
\label{4.6}
\ee
in order that  one can construct the one-form
\be
h^{-1} D h =  h^{-1} ( d + A ) h \equiv A'
\label{4.6a}
\ee
invariant under the $g$ transformation. Proceeding as before, 
the complete action can be determined  as 
\be
 S_{tot} = \int_{\cal M} \tr [ B \wed F] - \int_{\partial{\cal M}} 
\tr 
[ B \wed A' + \lambda (A'  \wed ^{*} A' ) ]
\label{4.7}
\ee
The constraint that follows by varying $A$ is given by 
\be
\lambda h^{-1} Dh = 0.
\label{4.8}
\ee 
while the equation of motion for the
 $h$ field turns out to be 
\be
d(*\lambda ( h^{-1} D h) -   B)= 0 .
\label{4.9}
\ee
Compatibility of (\ref{4.8},\ref{4.9}) requires 
\be
d B = 0
\label{4.10}
\ee
on the boundary.
as in the abelian case. However, this boundary equation is not invariant
under the transformation (\ref{4.5a}). 
 
\section{An Alternative Action in Four Dimensions}

The construction shown in previous sections suffers from an aesthetically 
unpleasant feature, namely the introduction of the Hodge $*$ operation
which requires the definition of a metric on the boundary.
This  prevents
the field theory on the boundary from being a``topological" theory. 
However, when we are working in four dimensional bulk
spacetime so that the boundary happens to be a three dimensional 
surface, one can write a Chern-Simons action on the boundary which does not 
require any metric. So,
the complete theory is diffeomorphism  
invariant and a topological field theory as well! 

So, the complete action in the latter case would be given by
\be
S_{total} = \int_{\cal M} B \wed F - \int_{\partial {\cal M}} 
[ B \wed A' + (A' dA' +  
\frac{2}{3} A' \wed A' \wed A') ]
\label{5.1}
\ee
where $A' = h^{-1} D h $ as in eq. (\ref{4.6a}). 

As before we use also the boundary action , consisting of  
the terms defined on the 
boundary in the 
above equation, to get the equation by varying $A$, namely,

\be 
 F(A')|_{\partial{\cal M}}= 0.
\label{5.2}
\ee

\section{The treatment for Gravity}

It is well-known that the vacuum Einstein equations can  
be obtained from the variation of the first order  
Palatini action ( also known as Einstein-Cartan 
action for our case) given by 
\be
S_{grav} = \int_{\cal M} d^4x \epsilon_{abcd} \epsilon^{\mu 
\nu \rho \lambda} e_\mu^a e_\nu^b R_{\rho \lambda}^{cd}
\equiv \int_{\cal M} \tr ( e \wed e \wed R)
\label{6.1}
\ee
where $e_\mu^a$ are the components of the tetrad and $a,b,..$ are  
the indices indicating that the object is valued in the $SO(3,1)$ 
Lie algebra. The curvature tensor is obtained from the spin connection:
\be
R_{ab} = d \omega_{ab} + \omega_a^c \wed \omega_{cb}
\label{6.2}
\ee
To obtain the equation of motion, one varies the tetrad and the  
spin connection independently. Note that 
this action is very similar to the   
non-abelian BF action, the important 
difference being  that one has the two-form 
$B$ replaced by the two-form $e \wedge e$ constructed 
out of the one-form $e$. So, one does not have the 
extra gauge invariance (\ref{2.10}) for the $B$ field anymore.
However, this does not deter us from proceeding 
in a fashion similar to the BF case using the variational arguments.
 In fact there exist formulations  
of gravity \cite{CDJ,hooft}
where one replaces the Einstein-Cartan action by a BF like 
action augemented by an extra term which basically enforces the algebraic
condition $B = e\wed e$. These extra terms do not involve the  
spin connection.
One sees that the analog of (\ref{2.4c})
now is
\be
\del e |_{\partial {\cal M}}=0
\label{6.3}
\ee
which tells us that we are fixing the metric on the boundary. This boundary 
can be the (stretched) horizon of the black hole.  

Varying $\omega$, we find the boundary term 

\be
\int_{\partial {\cal M}} E \wedge E \wedge  \del \Omega ,
\label{6.3.5}
\ee
where $E$ and  $\Omega$ are the pullbacks of
the tetrad $e$ and the spin connection 
$\omega$ respectively to the boundary $\partial M$.
 Therefore, the term to be added to the 
action is
\be
\int_{\partial {\cal M}} \epsilon_{abcd} E^a \wedge E^b \wedge 
\Omega^{cd}
\label{6.4}
\ee

Like for gauge invariance in the
BF case, the boundary term is not invariant under the local Lorentz 
transformation
\bea
\Omega \ra \Lambda \Omega \Lambda^T + \Lambda d \Lambda^T
\label{6.5}
\eea
where $\Lambda$ is valued in the SO(3,1) group (in its $4 \times 4$ 
irreducible representation), so 
$\Lambda_\mu^\alpha \Lambda_\nu^\beta \eta_{\
\alpha \beta}  
= \eta_{\mu \nu}$. This 
situation can be remedied by  replacing  the connection form $\Omega$ with the 
one-form 
\be
\Omega' \equiv u^T ( d + \Omega) u 
\label{6.6}
\ee
which is invariant under a local Lorentz transformation  
provided $u$ is in the same representation 
as $\Lambda$ and transforms as 
\be
u \ra \Lambda u 
\label{6.6a}
\ee
under this transformation. Note that the
 above construction is absolutely 
parallel to the construction  (\ref{4.6a})
in  the  non-abelian BF theory. One can see that the modified action is
\be
S_{mod.} = \int_{\cal M} \tr ( e \wed e \wed R) + \int_{\partial {\cal M}}
 E \wedge E \wedge \Omega'.
\label{6.7}
\ee

The boundary term in above action has appeared in a different context.
This type of action describes the coupling of 
 extended objects to external gauge fields \cite{al}. This boundary action 
would then describe the coupling of a membrane to a non-abelian connection.
It may be that a stertched
membrane description\cite{paradigm,maggiore} arises
naturally from 
this ``gauge degrees'' on the boundary. In fact, a similar picture 
emerges in the case of the 2+1 dimensional black hole \cite{carlip}.

In this treatment,  we have fixed 
$e$  and hence 
the metric on the boundary, and this could be a black hole metric too.
It is interesting to note that then this shows that there are gauge degrees 
of freedom living on the boundary of the black hole in the same sense 
as in \cite{us,carlip}. 
  
One can also add a Chern-Simons action constructed
from the one-form $\Omega'$ , as in section V, to the  action
(\ref{6.7}).
The constraint one gets by varying $\Omega'$  is 
\be
 R (\Omega')|_{\partial M} = 0.
\label{6.8}
\ee

This boundary condition is interesting. This tells us that one is 
really interested in the flat $SO(3,1)$ connections on the boundary.
However, the boundary of spacetime has the topology $\real \times
\Sigma $ where $\Sigma$ is a surface surrounding the black hole. 

The moduli space of the flat connections valued in the Lie algebra 
of compact groups are known to be finite dimensional. Though
$SO(3,1)$ is not  compact, a compact subgroup of it is all that 
we might be interested in and this might of relevance to the finiteness of  
the black hole entropy.

\section{Conclusions}

In this note, we have derived an action for the edge variables whose 
bulk dynamics is given by a four-dimensional BF theory. Note that in our 
approach it is not necessary that the field theory in the bulk is given by a
pure BF theory. One can add terms like  the integral of
$ B \wed B$ 
to the bulk action.
So, the above analysis can be applied very easily to the 
first order formalisms
for gravity. In fact, one notices that in gravity the two-form $B$ can be 
thought of as $e \wed e$ where $e$ is the tetrad one-form. Hence, naively 
speaking, the boundary condition for  $B$ fixes the 
metric on the boundary ( which can be null as well). 
The boundary term needed to make the variation well-defined 
is not invariant under the local SO(3,1) transformations
showing that there are gauge degrees of freedom on the 
boundary. 

It would be instructive to see whether the above states give rise to 
an entropy with interesting properties
like their three dimensional counterpart \cite{carlip}.

\section*{Acknowledgments}

I would like to thank my advisor 
Prof. A.P. Balachandran for his valuable guidance, support 
and constant encouragement.
I would also
like to thank J. Goldberg, T.R. Govindarajan, 
D. Marolf, V. John and  S. Vaidya 
for discussions. I am specially grateful to L. Chandar who
participated in the  initial stages of this work. 
I am indepted to 
Steve Carlip, S. Major  and Siddharta Sen for  
providing valuable criticisms on earlier 
versions of this paper. Part of this work was completed 
during TASI'96 and I would like to the organizers for 
their hospitality. This work was supported in 
part by the US Department of Energy under the 
contract number DE-FG02-85ER40231.

\end{document}